\documentclass[a4paper,12pt]{article}
\usepackage{fullpage}
\usepackage{alltt}
\title{CoInduction in Coq}
\author{Yves Bertot}
\newcommand{\coqand}{/\char'134}

\begin{document}
\hfuzz=3pt
\maketitle

When providing a collection of constructors to define an inductive type,
we actually also define a dual operation: a destructor.  This destructor
is always defined using the same structure of pattern-matching, so that
we have a tendency to forget that we do extend the ``pattern-matching''
capability with a new destructor at each definition.

Constructors and destructors play a dual role in the definition of
inductive types. Constructors produce elements of the inductive type,
destructors consume elements of the inductive type.

The inductive type itself is defined as the smallest collection of elements
that is stable with respect to the constructors: it must contain all constants
that are declared to be in the inductive type and all results of the
constructors when the arguments of these constructors are already found to be
in the inductive type.  When considering structural recursion,
recursive definitions are functions that consume elements of the inductive
type.  The discipline of structural recursion imposes that recursive calls
{\em consume} data that is {\em obtained} through the destructor.

The inductive type uses the constructors and destructors in a
specific way.  Co-inductive types are the types one obtains when using them
in a dual fashion.   A co-inductive type will appear as the largest
collection of elements that is stable with respect to the destructor.  It
contains every object that can be destructed by pattern-matching.

The duality goes on when considering the definition of recursive
functions.  Co-recursive functions are function that produce elements
of the co-inductive type.  The discipline of guarded co-recursion
imposes that co-recursive calls {\em produce} data that is {\em
consumed} by a constructor.  The main practical consequence is that
co-inductive types contain objects that look like {\em infinite
objects}.

This rough sketch is more of a philosophical nature.  When looking at the
details, there are some aspects of co-inductive types that are not so
simple to derive from a mere reflection of what happens with inductive types.

The possibility to have co-inductive types in theorem proving tools was
studied by Coquand \cite{coquand93}, Paulson \cite{paulson94},
Leclerc and Paulin-Mohring \cite{LeclercPaulin93}, and Gimenez 
\cite{gimenez94}.  Most of these authors were inspired by Aczel \cite{Aczel88}.
The paper \cite{Aczel02} provides a short presentation of terms and (possibly
infinite) trees, mainly in set-theoretic terms; it also
explains recursion and co-recursion.

In this document, we only consider the use of co-inductive types as it is
provided in Coq.
\section{Defining a co-inductive types}
Since defining a set of constructors automatically defines a destructor, the
definition of co-inductive types also relies on the definition of constructors.
The same rules of positivity as for inductive types apply.  Here are three
simple examples of co-inductive types:

\begin{alltt}
CoInductive Llist (A:Set) : Set :=
  Lcons : A -> Llist A -> Llist A | Lnil : Llist A.

CoInductive stream (A:Set) : Set :=
  Cons : A -> stream A -> stream A.

CoInductive Ltree (A:Set) : Set :=
  Lnode : A -> Ltree A -> Ltree A -> Ltree A  | Lleaf : Ltree A.
\end{alltt}

As for inductive types, this defines the type and the constructors, it also
defines the destructor, so that every element of the co-inductive can be
analysed by pattern-matching.  However, the definition does not provide
an induction principle.  The reason for the absence of an induction principle
can be explained in two ways, philosophical or technical.  Philosophically,
the induction principle of an inductive type expresses that this inductive
type is minimal (it is a least fixed point), but the co-inductive type
is rather viewed as greatest fixed point.  Technically, the induction
principle actually is a {\em consumption} tool, that consumes elements of
an inductive type to produce proofs in some other type, programed by recursion.
However, a co-recursive function can only be used to produce elements of
the co-recursive type, so that the only way to deduce anything from an
element of a co-inductive type is by pattern-matching.

The type {\tt Llist} given above comes with constructors {\tt Lcons} and
{\tt Lnil}.  These constructors will make it possible to produce lists,
exactly like the lists that we could produce in the inductive type {\tt list}.
Here are a few examples.
\begin{alltt}
Implicit Arguments Lcons.
Implicit Arguments Cons.

Definition ll123 := Lcons 1 (Lcons 2 (Lcons 3 (Lnil nat))).

Require Import List.

Fixpoint list_to_llist (A:Set) (l:list A)
    \{struct l\} : Llist A :=
 match l with
   nil => Lnil A
 | a::tl => Lcons a (list_to_llist A tl)
 end.

Definition ll123' := list_to_llist nat (1::2::3::nil).
\end{alltt}
The function {\tt list\_to\_llist} is recursive and produces elements in
a co-inductive type, but we did not make it rely on co-recursion.  Rather,
it relies on structural recursion as it is provided with the inductive
type {\tt list}.

Similar examples which do not rely on co-recursion cannot be provided for
the type {\tt stream}, because elements of this type cannot be given with
a finite number of uses of the constructor.  There is always a need for
another element of the co-inductive type.  Co-recursion is the solution.

In the coq system, co-recursion is only allowed under a form that can
ensure the strong normalization properties that are already satisfied by
the inductive part of the calculus.  The decision was taken to impose
a syntactic criterion: co-recursive values can only appear as argument
to constructors and inside branches of pattern-matching constructs.  Here
is a simple example:
\begin{alltt}
CoFixpoint ones : stream nat := Cons 1 ones.
\end{alltt}

This definition underlines a funny aspect of co-recursion: a
co-recursive value is not necessarily a function, because it is only
constrained to {\em produce} an element of the co-inductive type.  The
definition contains a usage of the constructor {\tt Cons} and a
reference to the co-recursive value itself, but this co-recursive
value is used as an argument to the constructor.

A similar value can be defined in the type {\tt Llist}.
\begin{alltt}
CoFixpoint lones : Llist nat := Lcons 1 lones.
\end{alltt}
Clearly, the list that we obtain is not a list that we could have obtained
using the function {\tt list\_to\_llist}.  The type {\tt Llist} is ``larger''
than the type {\tt list}.

Some co-recursive functions can be defined to perform exactly like similar
recursive functions on inductive types.  Here is an instance:
\begin{alltt}
Fixpoint map (A B:Set)(f:A -> B)(l:list A) \{struct l\} : list B :=
  match l with 
    nil => nil
  | a::tl => f a::map A B f tl
  end.

CoFixpoint lmap (A B:Set)(f:A -> B)(l:Llist A) : Llist B :=
  match l with
    Lnil => Lnil B
  | Lcons a tl => Lcons (f a) (lmap A B f tl)
  end.
\end{alltt}
The two functions look similar, but we should bear in mind that the second one
can also process infinite lists like {\tt lones}.

\section{Computing with co-recursive values}
When we manipulate elements of inductive types, we implicitely expect to
look at these values in constructor form: constructors applied to other
terms in constructor form.  However, attempting to put a value like
{\tt lones} in constructor form would require an infinity of unfoldings
of its value, this would make the computation non-normalizing.  For this
reason, a co-recursive value is by default considered to be a normal form.
This can be verified by requesting a computation on such a value.
\begin{alltt}
Eval simpl in lones.
\textit{ = lones : Llist nat}
\end{alltt}
However destructing a co-recursive value (with the help of a pattern-matching
construct) corresponds to a regular redex and
we can check that the first element of {\tt lones} indeed is 1.
\begin{alltt}
Eval simpl in
 match lones with Lnil => 0 | Lcons a _ => a end.
\textit{ = 1 : nat}
\end{alltt}
\section{Proving properties of co-inductive values}
It is possible to prove that two co-inductive values are equal.  The
usual approach, identical to what happens in the inductive setting is
to show that the two values have the same head constructor applied to
the same arguments.  However, making the head constructor appear is tricky,
because the computation of co-recursive values is usually not
performed.  One way to provoke this computation is to rely on the
following function and theorem.
\begin{alltt}
Definition Llist_decompose (A:Set)(l:Llist A) : Llist A :=
  match l with Lnil => Lnil A | Lcons a tl => Lcons a tl end.
Implicit Arguments Llist_decompose.

Theorem Llist_dec_thm :
   forall (A:Set)(l:Llist A), l = Llist_decompose l.
Proof.
 intros A l; case l; simpl; trivial.
Qed.
\end{alltt}
Now, here is an example using this theorem:
\begin{alltt}
Theorem lones_cons : lones = Lcons 1 lones.
Proof.
pattern lones at 1; rewrite Llist_dec_thm; simpl.
\textit{...
  ============================
   Lcons 1 lones = Lcons 1 lones
}
trivial.
Qed.
\end{alltt}
This is not a proof by co-recursion, just a proof by pattern-matching.

There are proofs of equality that seem obvious but cannot be performed
in the calculus of inductive constructions as it is defined now.  This
happens when the proofs seems to require some sort of inductive argument.
Here is an instance of an impossible proof:
\begin{alltt}
Theorem lmap_id : forall (A:Set)(l:Llist A),
   lmap A A (fun x:A => x) l = l.
\end{alltt}
One would like to have an argument of the following form: if the list
is nil, then the proof is trivial, if the list is not {\tt nil}, then
the head on both sides of the equality are naturally the same, and the
equality for the tails should hold by some ``inductive'' argument.  However,
there is no induction hypothesis, because there is no inductive list in
this statement and the proof of equality can only be proved by using
the constructor of equality (because equality is itself an inductive type).

The solution for this kind of problem is to use a co-inductive proposition
that expresses the equality of two lists by stating that they have the
same elements.  Here is the co-inductive definition for this proposition:
\begin{alltt}
CoInductive bisimilar (A:Set) : Llist A -> Llist A -> Prop :=
  bisim0 : bisimilar A (Lnil A) (Lnil A)
| bisim1 : forall a l l', 
       bisimilar A l l' -> bisimilar A (Lcons a l)(Lcons a l').
\end{alltt}
Proofs that two lists have the same elements can now also be obtained
by using co-recursive values, as long as we use the {\tt bisimilar}
relation instead of equality.  Here is an example of a proof, displayed
as term of the calculus of inductive constructions to make the general
structure visible.  Please note that rewrites using the theorems
{\tt eq\_ind\_r} and {\tt Llist\_dec\_thm} are performed to introduce
the function {\tt Llist\_decompose} and force the expansion of the
co-recursive function.

\begin{alltt}
CoFixpoint lmap_bi (A:Set)(l:Llist A) :
  bisimilar A (lmap A A (fun x:A => x) l) l :=
 @eq_ind_r (Llist A) (Llist_decompose (lmap A A (fun x=> x) l))
   (fun x => bisimilar A x l)
   match l return bisimilar A 
                    (Llist_decompose (lmap A A (fun x=>x) l))
                    l with
     Lnil => bisim0 A
   | Lcons a k =>
             bisim1 A a (lmap A A (fun x=> x) k) k (lmap_bi A k)
   end
   (lmap A A (fun x => x) l)
   (Llist_dec_thm A (lmap A A (fun x=>x) l))
.

\end{alltt}
The manual construction of co-inductive proofs is difficult.  The
alternative approach is to use tactics.  The following script performs
the same proof, but relying on tactics.
\begin{alltt}
Theorem lmap_bi' : forall (A:Set)(l:Llist A),
  bisimilar A (lmap A A (fun x => x) l) l.
cofix.
intros A l; rewrite (Llist_dec_thm _ (lmap A A (fun x=>x) l)).
case l.
intros a k; simpl.
apply bisim1; apply lmap_bi'.
simpl; apply bisim0.
Qed.
\end{alltt}
The tactic {\tt cofix} is the tactic that declares that the current proof
will be a co-recursive value.  It introduces a new assumption in the context
so that the co-recursive value can be used inside its own definition.
However, the same constraints as before exist: the co-recursive value can
only be used as input to a constructor.  In the case of {\tt lmap\_bi},
the use of {\tt lmap\_bi'} at the end of the proof is justified by the
previous use of the constructor {\tt bisim1}: {\tt lmap\_bi'} is thus
used to provide an argument to {\tt bisim1}.

In general, the constraint that co-recursive calls are used in correct
conditions is only checked at the end of the proof.  This sometimes has
the unpleasant effect that one believes to have completed a proof and
is only rebuked when the {\tt Qed} or {\tt Defined} commands announce that
the constructed term is not well-formed.  This problem is compounded by
the fact that it is hard to control the hypotheses that are used by automatic
tactics.  Even though we believe the proof of a subgoal should not rely
on the co-recursive assumption, it may happen that some tactic like
{\tt intuition} uses this assumption in a bad way.  One solution to this
problem is to use the {\tt clear} tactic to remove the co-recursive assumption
before using strong automatic tactics.  A second important tool to avoid
this problem is a command called {\tt Guarded}, this command can be used
at any time during the interactive proofs and it checks whether illegal
uses of the co-recursive tactic have already been performed.
\section{Applications}
Co-inductive types can be used to reason about hardware descriptions
\cite{coupet3} concurrent programming \cite{gimenez95}, finite
state automata and infinite traces of execution, and temporal logic
\cite{dr1,coupet2}.  The
guarded by constructors structure of co-recursive functions is adapted
to representing finite state automata.  A few concrete examples
are also given in \cite{coqart}.

Co-inductive types are especially well suited to model and reason
about lazy functional programs that compute on infinite lists.
However, the constraints of having co-recursive calls guarded by
constructors imposes that one scrutinizes the structure of recursive
functions to understand whether they really can be encoded in the
language.  One approach, used in \cite{filters05} is to show that
co-inductive objects also satisfy some inductive properties, which
make it possible to define functions that have a recursive part, with
usual structural recursive calls with respect to these inductive
properties, and guarded co-recursive parts.

\section{An example: introduction to exact real arithmetics}
The work presented in this section is my own, but it is greatly inspired
by reading the lecture notes \cite{EdalatHeckmann02} and the thesis
\cite{niqui04} and derived papers \cite{niqui05,niqui05b}, and by
\cite{CiaffaglioneDiGianantonio99}.  These
papers should be consulted for further references about exact real
arithmetics, lazy computation, and co-inductive types.

We are going to represent real numbers between 0 and 1 (included) as
infinite sequences of intervals
\(I_n\), where \(I_0=[0,1]\), \(I_{n+1}\subset I_{n}\) and the size
of \(I_{n+1}\) is
half the size of \(I_{n}\).  Moreover, \(I_{n+1}\) will be obtained from
\(I_{n}\) in only one of three possible ways:
\begin{enumerate}
\item \(I_{n+1}\) is the left half of \(I_{n}\),
\item \(I_{n+1}\) is the right half of \(I_{n+1}\),
\item \(I_{n+1}\) is the center of \(I_{n+1}\): if \(a\) and \(b\) are the
bounds of \(I_n\), then \(a+(b-a)/4\) and \(a+3(b-a)/4\) are the bounds of
\(I_n+1\).
\end{enumerate}
We can represent any of the intervals \(I_n\) using lists of {\tt
idigit}, where the type {\tt idigit} is the three element enumerated
type containing {\tt L}, {\tt R}, {\tt C}.  For instance, the
interval [0,1] is given by the empty list, the interval [1/4,3/8] can
be represented by the lists {\tt L::C::R::nil}, {\tt L::R::L::nil}, or
{\tt C::L::L::nil}.  It is fairly easy to write a function of type {\tt
list idigit->R} that maps every list to the lower and upper bound of
the interval it represents.  We are going to represent real numbers by
infinite sequences of intervals using the type {\tt stream idigit}.

There is also an easy correspondence from floating-point numbers in
 binary representation to this representation.  Let us first recall what
the binary floating-point representation is.  Any ``binary'' floating point
is a list of boolean values.  Interpreting {\tt true} as the 1 bit
and {\tt false} as the 0 bit, a boolean list is interpreted as a real number
in the following way:
\begin{alltt}
Fixpoint bit_list_to_R (l:list boolean) : Rdefinitions.R :=
  match l with 
    nil => 0
  | b::tl => let x := bit_list_to_R tl in
             if b then (1+x)/2 else x/2
  end.
\end{alltt}
We can inject the boolean values into the type {\tt idigit} mapping 
{\tt true} to {\tt L} and {\tt false} to {\tt R}.  It is fairly easy
to show that this correspondance can be lifted to lists of booleans and
idigits, so that the real number represented by a list is element of
the interval represented by the corresponding list.

We represent real numbers by streams of {\tt idigit} elements.  The
construction relies on associating a sequence of real numbers
to each stream (actually the lower bounds of the intervals) and to
show that this sequence converges to a limit.  To ease our reasoning,
we will also describe the relation between a stream and a real value
using a co-inductive property:
\begin{alltt}
CoInductive represents : stream idigit -> Rdefinitions.R -> Prop :=
  reprL : forall s r, represents s r -> (0 <= r <= 1)\%R ->
           represents (Cons L s) (r/2)
| reprR : forall s r, represents s r -> (0 <= r <= 1)\%R ->
           represents (Cons R s) ((r+1)/2)
| reprC : forall s r, represents s r -> (0 <= r <= 1)\%R ->
           represents (Cons C s) ((2*r+1)/4).
\end{alltt}

We could also use infinite lists of booleans to represent real
numbers.  This is the usual representation of numbers.  This
representation also corresponds to
sequences of intervals, but it has bad programming
properties.  In this representation, if we know that a number is
very close to 1/2 but we don't know whether it is larger or smaller, we
cannot produce the first bit.  For instance,
the number 1/3 is represented by the
infinite sequence .0101\dots and the number 1/6 is represented by the infinite
sequence .0010101\dots  Adding the two numbers should yield the number \(1/2\).
However, every finite prefix of 
.010101\dots represents an interval that contains numbers that
are larger than \(1/3\) and numbers that are smaller than \(1/3\).  Similarly,
every finite prefix of .0010101\dots contains a numbers that are larger
than \(1/6\) and numbers that are smaller.  By only looking at a finite
prefix of both numbers, we cannot decide whether the first bit of the
result should be a 0 or a 1, because no number larger than \(1/2\) can be
represented
by a sequence starting with a 0 and no number smaller than \(1/2\) can
be represented by a sequence starting with a 1.

With the extra digit, {\tt C}, we can perform the computation as follows:
\begin{enumerate}
\item having observed that the first number has the form \(x=LRx'\),
we know that this number is between 1/4 and 1/2,
\item having observed that the second number has the form \(y=LLy'\), we
know that this number is between 0 and 1/4,
\item we know that the sum is between 1/4 and 3/4.  therefore, we know
that the sum is an element of the interval represented by {\tt C::nil}, and we
can output this digit.
\end{enumerate}

We can also go on to output the following digits.  In usual binary
representation, if \(v\) is the number represented by the sequence
\(s\), then the number represented by the sequence \(0s\) is \(v/2\)
and the number represented by the sequence \(1s\) is \((v+1)/2\).
This interpretation carries over to the digits {\tt L} and {\tt R},
respectively.  For the digit {\tt C}, we know that the sequence {\tt
C\(s\)} represents \((2v+1)/4\).  Thus, if we come back to the
computation of
\(1/3+1/6\), we know that \(x'\) is \(4*x-1\), \(y'\) is \(4*y\), and
the result should have the form {\tt C::\(z\)}, where \(z\) is the
representation of
\((x'+y'+1)/4\) (since \(x'+y'+1)/4\) is 1/2, we see that the result
of the sum is going to be an infinite sequence of {\tt C} digits.

We are now going to provide a few functions on streams.
As a first example, the function {\tt rat\_to\_stream} maps any
two integers \(a\) \(b\) to a stream.  When \(a/b\) is between 0 and
1, the result stream is the representation of this rational number.

\begin{alltt}
CoFixpoint rat_to_stream (a b:Z) : stream idigit :=
  if Z_le_gt_dec (2*a) b then
    Cons L (rat_to_stream (2*a) b)
  else
    Cons R (rat_to_stream (2*a-b) b)
\end{alltt}
For the second example, we compute an affine combination
of two numbers with rational coefficients.  We will define the function
that constructs the representation of the following formula.
\[\frac{a}{a'}v_1 + \frac{b}{b'}v_2 + \frac{c}{c'}\]
The numbers \(a\), \(a'\), \dots are positive integers and \(a'\), \(b'\),
and \(c'\) are non-zero (this sign restriction only serves to make the
example shorter).

We choose to define a one-argument function, where the argument
is a record holding all the values \(a\), \(a'\), \dots, \(v_1\), \(v_2\).
We define a type for this record and a predicate
to express the sign conditions.
\begin{alltt}
Record affine_data : Set :=
 \{m_a : Z; m_a' : Z; m_b : Z; m_b' : Z; m_c : Z; m_c' : Z;
  m_v1 : stream idigit; m_v2 : stream idigit\}.

Definition positive_coefficients (x:affine_data) :=
  0 <= m_a x \coqand{} 0 < m_a' x \coqand{} 0 <= m_b x \coqand{} 0 < m_b' x
  \coqand{} 0 <= m_c x \coqand{} 0 < m_c' x.
\end{alltt}
We define a function {\tt axbyc} of type
\begin{alltt}
 forall x, positive_coefficients x -> stream idigit.
\end{alltt}
The algorithm contains two categories
of computing steps.  In computing steps of the first category, a digit
of type {\tt idigit} is produced, because analysing the values of
\(a\), \(a'\), \dots makes it possible to infer that the result will be
in a precise part of the interval.  The result then takes the form
\begin{center}
{\tt Cons \(d\) (axbyc \(\langle a_1,a'_1,b_1,b'_1,c_1,c'_1,v_1,v_2\rangle\))}
\end{center}
Where \(d\) is a digit and the values of \(a_1\), \(a'_1\), \dots
depend on the digit.
\begin{enumerate}
\item if \(c/c' \geq 1/2\), then the result is sure to be in the right
part of the interval, the digit \(d\) is {\tt R} and
the new parameters are chosen so that
\(a_1/a'_1= 2a/a'\), \(b_1/b'_1= 2b/b'\), \(c_1/c'_1=(2c-c')/c'\), because
of the following equality:
\[\frac{1}{2}(\frac{2a}{a'}v_1 +\frac{2b}{b'} v_2 +
\frac{2c-c'}{c'})+\frac{1}{2}=
\frac{a}{a'}v_1+\frac{b}{b'}v_2+\frac{c}{c'}\]
\item if \(2(ab'c'+ba'c'+a'b'c)\leq a'b'c'\),
      then the result is sure to be in the
left part of the interval, the digit \(d\) is {\tt L} and
the new parameters are chosen so that
 \(a_1/a'_1= 2a/a'\), \(b_1/b'_1= 2b/b'\), \(c_1/c'_1=2c/c'\) (we
do not detail the justification),
\item if \((4(ab'c'+ba'c'+a'b'c) \leq 3a'b'c'\) and \(4*c\geq c'\), then the
result is sure to belong to the center sub-interval, the digit \(d\) is
{\tt C} and the new parameters are
chosen so that
 \(a_1/a'_1= 2a/a'\), \(b_1/b'_1= 2b/b'\), \(c_1/c'_1=(4c-c')/2c'\).
\end{enumerate}
The various cases of these productive steps are described using the
following functions:
\begin{alltt}
Definition prod_R x :=
  Build_affine_data (2*m_a x) (m_a' x) (2*m_b x) (m_b' x)
  (2*m_c x - m_c' x) (m_c' x) (m_v1 x) (m_v2 x).

Definition prod_L x :=
  Build_affine_data (2*m_a x) (m_a' x) (2*m_b x) (m_b' x)
  (2*m_c x) (m_c' x) (m_v1 x) (m_v2 x).

Definition prod_C x :=
  Build_affine_data (2*m_a x) (m_a' x) (2*m_b x) (m_b' x)
  (4*m_c x - m_c' x) (2*m_c' x) (m_v1 x) (m_v2 x).
\end{alltt}

In the second category of computing steps the values \(v_1\)
and \(v_2\) are scrutinized, so that the interval for the potential values
of the result is reduced as one learns more information about the inputs.
If the values \(v_1\) and \(v_2\) have the form {\tt Cons \(d_1\) \(v'_1\)}
and {\tt Cons \(d_2\) \(v'_2\)} respectively, The result then takes the form
\begin{center}
 {\tt axbyc \(\langle a,2a',b,2b',c_1,c'_1,v'_1,v'_2\rangle\)}
\end{center}
Only the parameters \(c_1\) and \(c'_1\) take a different form depending on
the values of \(d_1\) and \(d_2\).  The correspondance is given
in the following table.
\[\begin{array}{llrr}
d_1&d_2&c_1&c'_1\\
\hline
{\tt L}&{\tt L}&c&c'\\
{\tt L}&{\tt R}&bc'+2cb'&2b'c'\\
{\tt R}&{\tt L}&ac'+2ca'&2a'c'\\
{\tt L}&{\tt C}&bc'+4cb'&4b'c'\\
{\tt C}&{\tt L}&ac'+4ca'&4a'c'\\
{\tt R}&{\tt C}&2ba'c'+ab'c'+4cb'a'&4a'b'c'\\
{\tt C}&{\tt R}&2ba'c'+ab'c'+4cb'a'&4b'a'c'\\
{\tt R}&{\tt R}&ab'c'+ba'c'+2ca'b'&2a'b'c'\\
{\tt C}&{\tt C}&ba'c'+ab'c'+4cb'a'&4b'a'c'
\end{array}
\]
For justification, let us look only at the case where \(v_1={\tt R}v'_1\) and
\(v_2={\tt C}v'_2\).  In this case we have the following equations:
\begin{eqnarray*}
\frac{a}{a'}v_1+\frac{b}{b'}v_2+\frac{c}{c'}&=&\frac{a}{a'}
(\frac{1}{2}v'_1+\frac{1}{2})+\frac{b}{b'}(\frac{1}{2}v'_2+\frac{1}{4})+
\frac{c}{c'}\\
&=&\frac{a}{2a'}v'_1+\frac{b}{2b'}v'_2+\frac{2ba'c'+ab'c'+4cb'a'}{4a'b'c'}
\end{eqnarray*}
This category of computation is taken care of by a function with
the following form:
\begin{alltt}
Definition axbyc_consume (x:affine_data) :=
 let (a,a',b,b',c,c',v1,v2) := x in
 let (d1,v1') := v1 in let (d2,v2') := v2 in
 let (c1,c1') :=
  match d1,d2 with
  | L,L => (c, c')
  | L,R => (b*c'+2*c*b', 2*b'*c')
  | R,L => (a*c'+2*c*a', 2*a'*c')
  | L,C => (b*c'+4*c*b', 4*b'*c')
  | C,L => (a*c'+4*c*a', 4*a'*c')
  | R,C => (2*a*b'*c'+b*a'*c'+4*c*a'*b', 4*a'*b'*c')
  | C,R => (2*b*a'*c'+a*b'*c'+4*c*b'*a', 4*b'*a'*c')
  | R,R => (a*b'*c'+b*a'*c'+2*c*a'*b', 2*a'*b'*c')
  | C,C => (b*a'*c'+a*b'*c'+4*c*b'*a', 4*b'*a'*c')
  end in
 Build_affine_data a (2*a') b (2*b') c1 c1' v1' v2'.
\end{alltt}

From the point of view of co-recursive programming, the first category
of computing steps gives regular guarded-by-constructor corecursive
calls.  The second category of computing steps does not give any
guarded corecursion.  We need to separate the second category in an
auxiliary function.  We choose to define this auxiliary function by
well-founded induction.  The recursive function performs the various
tests with the help of an auxiliary test function:
\begin{alltt}
  Parameter axbyc_test :
  forall x,
   positive_coefficients x ->
   {m_c' x <= 2*m_c x}+
   {2*(m_a x*m_b' x*m_c' x +
        m_b x*m_a' x*m_c' x + m_a' x*m_b' x*m_c x)<=
     m_a' x*m_b' x*m_c' x}+
   {4*(m_a x*m_b' x*m_c' x +
        m_b x*m_a' x*m_c' x + m_a' x*m_b' x*m_c x)<=
     3*m_a' x*m_b' x*m_c' x \coqand{} m_c' x <= 4*m_c x}+
   {m_a' x < 8*m_a x \/ m_b' x < 8*m_b x}.
\end{alltt}
In the first three cases, the recursive function just returns
the value that it received, together with the proofs of the properties.
To carry these agregates of values and proofs, we defined
a specific type to combine these values and proofs.
\begin{alltt}
Inductive decision_data : Set :=
  caseR : forall x:affine_data, positive_coefficients x -> 
          m_c' x <= 2*m_c x -> decision_data
| caseL : forall x:affine_data, positive_coefficients x ->
          2*(m_a x*m_b' x*m_c' x +
             m_b x*m_a' x*m_c' x + m_a' x*m_b' x*m_c x)<=
           m_a' x*m_b' x*m_c' x -> decision_data
| caseC : forall x:affine_data, positive_coefficients x ->
          4*(m_a x*m_b' x*m_c' x +
             m_b x*m_a' x*m_c' x + m_a' x*m_b' x*m_c x)<=
          3*m_a' x*m_b' x*m_c' x -> m_c' x <= 4*m_c x ->
          decision_data.
\end{alltt}
The recursive function will thus have the type
\begin{alltt}
 forall x, positive_coefficient x -> decision_data.
\end{alltt}
The definition has the following form:
\begin{alltt}
Definition axbyc_rec_aux (x:affine_data)
   : (forall y, order y x ->
         positive_coefficients y -> decision_data)->
     positive_coefficients x -> decision_data :=
  fun f Hp =>
  match A.axbyc_test x Hp with
    inleft (inleft (left H)) => caseR x Hp H
  | inleft (inleft (right H)) => caseL x Hp H
  | inleft (inright (conj H1 H2)) => caseC x Hp H1 H2
  | inright H =>
    f (axbyc_consume x)
      (A.axbyc_consume_decrease x Hp H)
      (A.axbyc_consume_pos x Hp)
  end.

Definition axbyc_rec :=
  well_founded_induction A.order_wf
  (fun x => positive_coefficients x -> decision_data)
  axbyc_rec_aux.
\end{alltt}
The definition of {\tt axbyc\_rec} of course relies on proofs to ensure
that {\tt axbyc\_consume} preserves the sign conditions and make
the measure decrease, we do not include these proofs in these notes.

The main co-recursive function relies on the auxiliary recursive function
to perform all the recursive calls that are not productive,
the value returned by the auxiliary function is suited to produce data and
co-recursive calls are then allowed.
\begin{alltt}
CoFixpoint axbyc (x:affine_data) 
   (h:positive_coefficients x):stream idigit :=
  match axbyc_rec x h with
    caseR y Hpos H => Cons R (axbyc (prod_R y) (A.prod_R_pos y Hpos H))
  | caseL y Hpos H => Cons L (axbyc (prod_L y) (A.prod_L_pos y Hpos))
  | caseC y Hpos H1 H2 => 
         Cons C (axbyc (prod_C y) (A.prod_C_pos y Hpos H2))
  end.
\end{alltt}
This function relies on auxiliary functions to perform the relevant
updates of the various coefficients.  For instance, here is the function
{\tt prod\_C}:
\begin{alltt}
Definition prod_C x :=
  Build_affine_data (2*m_a x) (m_a' x) (2*m_b x) (m_b' x)
  (4*m_c x-m_c' x) (m_c' x) (m_v1 x) (m_v2 x).
\end{alltt}
For each of these functions, it is also necessary to prove that they
preserve the sign conditions, these proofs are fairly trivial.

It requires more work to prove that the function is correct, in the
sense that it does produce the representation of the right real number,
but this proof is too long to fit in these short tutorial notes.  More
work is also required to make the function more efficient, for instance
by dividing {\tt a} (resp. {\tt b}, {\tt c}) and {\tt a'} (resp. {\tt b'},
{\tt c'}) by they greatest common divisor at each step.

The representation for real numbers proposed in
\cite{EdalatHeckmann02} is very close to the representation used in
these notes, except that the initial interval is [-1,1], and the three
digits are interpreted as the sub-intervals [-1,0], [0,-1],
[-1/2,1/2].  The whole set of real numbers is then encoded by
multiplying a number in [-1,1] by an exponent of 2 (as in usual
scientific, floating point notation).  The work presented in
\cite{niqui04} shows that both the representation in these notes
and the representation in \cite{EdalatHeckmann02} are a particular case
of a general framework based on overlapping intervals and proposes a
few other solutions.  In these notes, we have decided to restrict
ourselves to affine binary operations, which makes it possible to
obtain addition and multiplication by a rational number, but the most
general setting relies on homographic and quadratic functions, which
make it possible to obtain addition, multiplication, and division, all
in one shot.

The method of separating a recursive part from a co-recursive part in
a function definition was already present in \cite{filters05}.
However, the example of \cite{filters05} is more complex because the
functions are {\em partial}: there are streams for which eventual
productivity is not ensured and a stronger description technique is
required.  This stronger technique is described in \cite{coqart} as
{\em ad-hoc} recursion.  The papers
\cite{diGianantonioMiculan,diGianantonioMiculan04} propose an alternative
foundation to functions that mix recursive and co-recursive parts.

\section{Exercises}
\begin{description}
\item [increasing streams] Define a co-inductive predicate that is
satisfied by any stream such that, if \(n\) and \(m\) are consecutive
elements, then \(n \leq m\).
\item [Fibonnacci streams] Define a co-inductive predicate, called 
{\tt local\_fib}, that is
satisfied by any stream such that, if \(n\), \(m\), \(p\) are consecutive
elements, then \(p=n+m\).  Define a co-recursive function that 
constructs a fibonacci stream whose first two elements are 1.  Prove that
the stream that is created satisfies the two predicates ({\tt increasing}
and {\tt local\_fib}).
\end{description}
\pagebreak
\section{Solutions}
\begin{alltt}
Require Export Omega.

CoInductive increasing : stream nat -> Prop :=
  ci : forall a b tl, a <= b -> increasing (Cons b tl) ->
              increasing (Cons a (Cons b tl)).

CoInductive local_fib : stream nat -> Prop :=
  clf : forall a b tl, local_fib (Cons b (Cons (a+b) tl)) ->
       local_fib (Cons a (Cons b (Cons (a+b) tl))).

CoFixpoint fibo_str (a b:nat) : stream nat := Cons a (fibo_str b (a + b)).

Definition str_decompose (A:Set)(s:stream A) : stream A :=
  match s with Cons a tl => Cons a tl end.

Implicit Arguments str_decompose.

Theorem str_dec_thm : forall (A:Set)(s:stream A), str_decompose s = s.
Proof.
intros A [a tl];reflexivity.
Qed.

Implicit Arguments str_dec_thm.

Theorem increasing_fibo_str :
  forall a b, a <= b -> increasing (fibo_str a b).
Proof.
Cofix.
intros a b Hle.
rewrite <- (str_dec_thm (fibo_str a b));simpl
assert (Heq:(fibo_str b (a+b))=(Cons b (fibo_str (a+b) (b+(a+b))))).
rewrite <- (str_dec_thm (fibo_str b (a+b)));simpl;auto.
rewrite Heq.
constructor.
assumption.
rewrite <- Heq.
apply increasing_fibo_str.
omega.
Qed.

Theorem increasing_fib : increasing (fibo_str 1 1).
Proof.
 apply increasing_fibo_str;omega.
Qed.

Theorem local_fib_str :
  forall a b, local_fib (fibo_str a b).
Proof.
cofix.
intros a b.
assert (Heq :
          (fibo_str b (a+b)) =
          (Cons b (Cons (a+b)(fibo_str (b+(a+b))((a+b)+(b+(a+b))))))).
rewrite <- (str_dec_thm (fibo_str b (a+b))); simpl.
rewrite <- (str_dec_thm (fibo_str (a+b) (b+(a+b)))); simpl;auto.
rewrite <- (str_dec_thm (fibo_str a b)); simpl.
rewrite Heq.
constructor.
rewrite <- Heq.
apply local_fib_str.
Qed.

Theorem local_fib_fib : local_fib (fibo_str 1 1).
Proof.
 apply local_fib_str.
Qed.
\end{alltt}

\end{document}